\documentclass{article}
\usepackage{spconf,amsmath,graphicx}
\usepackage{subcaption}
\usepackage{spconf,amsmath,graphicx,amssymb,bm,textcomp,booktabs}
\usepackage{url}
\usepackage{subcaption}
\usepackage{siunitx}
\usepackage{cite}
\usepackage{caption}
\usepackage{tablefootnote}
\usepackage{tipa}
\usepackage{multirow}

\title{Into-TTS : Intonation Template Based Prosody Control System}
\name{Jihwan Lee, Joun Yeop Lee, Heejin Choi, Seongkyu Mun, Sangjun Park, Jae-Sung Bae, Chanwoo Kim}
\address{Samsung Research, Seoul, South Korea}
\begin{document}
%
\maketitle
\begin{abstract}
Intonations play an important role in delivering the intention of a speaker.
However, current end-to-end TTS systems often fail to model proper intonations.
To alleviate this problem, we propose a novel, intuitive method to synthesize speech in different intonations using predefined \textit{intonation templates}.
Prior to TTS model training, speech data are grouped into intonation templates in an unsupervised manner.
Two proposed modules are added to the end-to-end TTS framework: an \textit{intonation predictor} and an \textit{intonation encoder}.
The intonation predictor recommends a suitable intonation template to the given text.
The intonation encoder, attached to the text encoder output, synthesizes speech abiding the requested intonation template.
Main contributions of our paper are:
(a) an easy-to-use intonation control system covering a wide range of users;
(b) better performance in wrapping speech in a requested intonation with improved objective and subjective evaluation;
and (c) incorporating a pre-trained language model for intonation modelling.
Audio samples are available at \url{https://srtts.github.io/IntoTTS}.

\end{abstract}
\section{Introduction}
\label{sec:intro}

In oral conversations, it is crucial to deliver speech in an appropriate intonation to ensure the speaker's correct intention.
Despite its identical syntactic structure, a single sentence can convey various intentions with different intonations.
For example, in English, if a speaker asks a yes/no question in a falling intonation, then it is more probable that the speaker is demanding for confirmation or a gentle request rather than an actual answer, while the same interrogative sentence in a rising manner functions as a genuine question \cite{hudson1975meaning, hedberg2017meaning}.
Utterances wrapped in a poorly chosen intonation often lead to miscommunication. 

The same holds true for speech synthesis.
TTS systems should be able to synthesize speech in an appropriate intonation to the context, or at least allow users to choose one, in case they have a specific preference.
However, many of the existing TTS systems fail to do so, degrading its naturalness.
They usually take text as the only input.
With no additional cues being provided, they lack ability to synthesize a single sentence in various intonations.
To mitigate this issue, some approaches propose to control prosody by feeding auxiliary information along with the text.
One approach is utilizing a prosody embedding extracted from a reference encoder \cite{skerry2018towards}, followed by many other variations \cite{lee2019robust, wei2019ICLR, wei2019ICASSP, battenberg2020effective}.
Their methods allow users to control some prosody factors in detail.
However, to our best knowledge, they all require a reference audio for synthesis, leaving it a user's burden to explore.

The Global Style Tokens (GST) \cite{wang2018style} method utilizes tokens as soft labels to control speech styles.
Users can either feed a reference audio or adjust the tokens to synthesize speech in various styles.
However, the exact roles of each token are still ambiguous.
This does not guarantee that a specific style factor, such as intonation, can be properly reflected to the model.
Moreover, additional works should be preceded to investigate appropriate weights and tokens to map their numerical values to the corresponding style labels.
These reference encoder-based methods mainly aim to imitate the general style of the reference speech.
Hence, controlling a specific factor, such as intonation, is not a suitable task for these approaches.


In this work, we propose \textit{Into-TTS}, a novel, intuitive method to synthesize speech in different intonations using predefined Intonation Templates (IT).
We extend the template-based approaches \cite{ronanki2016template, vioni2021prosodic} to the intonation level, introducing intonation templates that are created using an unsupervised clustering method.
These predefined intonation templates provide users a simple solution to synthesize speech in an intonation pattern they demand, offering only a limited number of options to the users.
Moreover, the unsupervised clustering approach requires no additional annotation, therefore most of currently available speech corpora can be directly utilized.

In the proposed method, two modules are added to a Tacotron \cite{taco1, taco2} variant \cite{ellinas2020high}: an \textit{intonation predictor} and an \textit{intonation encoder}.
The intonation predictor learns to recommend an intonation template suitable to the given text.
The intonation encoder takes an intonation template index as input, yielding the corresponding intonation embedding to the decoder.
This enables the TTS model to synthesize speech in the given intonation accordingly.

Inference works in two different modes: AUTO and MANUAL.
In the AUTO mode, where no specific preference to an intonation pattern type is provided, the intonation predictor automatically recommends a suitable intonation to the text, then it's synthesized accordingly.
In the MANUAL mode, when a specific preference to an intonation is explicitly requested, the TTS model synthesizes speech in that specific intonation template.
In either mode, it is no longer necessary to control numerous factors individually, which allows users to easily utilize the system, even if they have little background knowledge.

Main contributions of this paper include but are not limited to:
\begin{itemize}
    \item A convenient intonation control system that can cover various types of users, even those without professional knowledge on TTS frameworks.
    \item Improved performance to synthesize speech in a desired intonation.
    
    \item Feasibility to incorporating pre-trained language model to intonation modelling in TTS frameworks. 
     
\end{itemize}

\section{Proposed Method}

The proposed method consists of two parts: (a) creating \textit{intonation templates} by unsupervised clustering; and (b) jointly training the TTS model with the \textit{intonation predictor} and the \textit{intonation encoder}, utilizing a predefined set of intonation templates.

\subsection{Intonation Templates}\label{into_temp}


Linguists typically classify English intonation into the following four patterns: \textit{declarative}, \textit{yes/no question}, \textit{incredulity}, and \textit{obviousness} \cite{liberman1975intonational}.
However, manually annotating these intonation patterns is not an easy task.
As many researchers have pointed out, it requires high cost and often yields unreliably annotated results, partly due to the complexity of the \textit{ToBI} \cite{silverman1992tobi} system \cite{wightman2002tobi,lee2019robust}.

In this paper, we propose a method to automatically classify intonations without any annotated audio data, creating intonation templates in an unsupervised manner.
First, all audio files are de-noised and trimmed with a very tight margin.
The trimming process minimizes false detection of F0 after the end of utterances, and eases sentence-final alignment among the recordings.
F0 is then extracted from speech by the RAPT \cite{talkin1995robust} algorithm.
The F0 values of unvoiced segments are linearly interpolated according to the F0 values of the nearby voiced segments.
Next, to eliminate the speaker variation effect on F0, the F0 values in each utterance are $L2$-normalized, i.e. dividing the F0 values by the square root of the sum of the squared F0 values.
After the normalization, the last 0.5 seconds of the F0 values from each recording are selected for clustering, since the mean length of sentence-final intonation segments in English is 0.37 seconds with the standard deviation of 0.15 \cite{berkovits1984duration}.
These sentence-final frames of F0 are automatically classified into $N$ types of intonations via k-means clustering.

\begin{figure}[t]
    \centering

    \begin{subfigure}[c]{.47\linewidth}
        \centering
        {\includegraphics[width=1.0\linewidth]{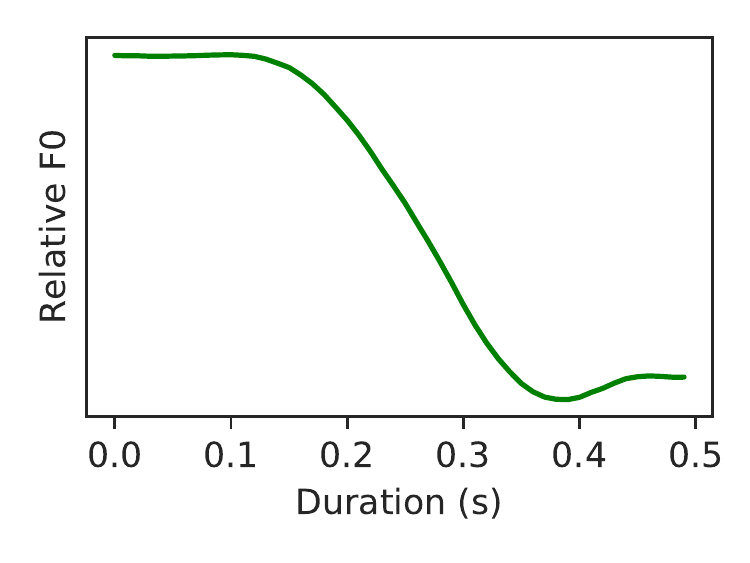}}
        \caption{Declarative}\medskip
        \label{fig:into_a}
    \end{subfigure}
    \hfill
    \begin{subfigure}[c]{.47\linewidth}
        \centering
        {\includegraphics[width=1.0\linewidth]{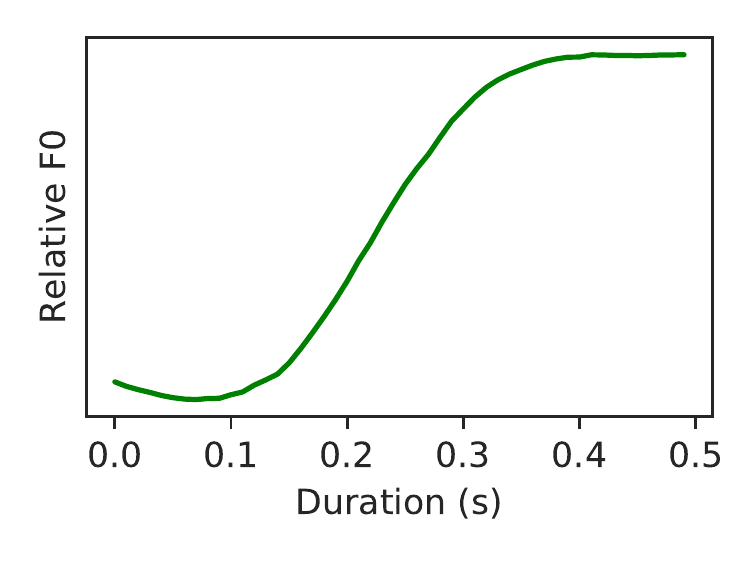}}
        \caption{Yes/no Question}\medskip
        \label{fig:into_b}
    \end{subfigure}
    \newline
    
    \begin{subfigure}[c]{.47\linewidth}
        \centering
        {\includegraphics[width=1.0\linewidth]{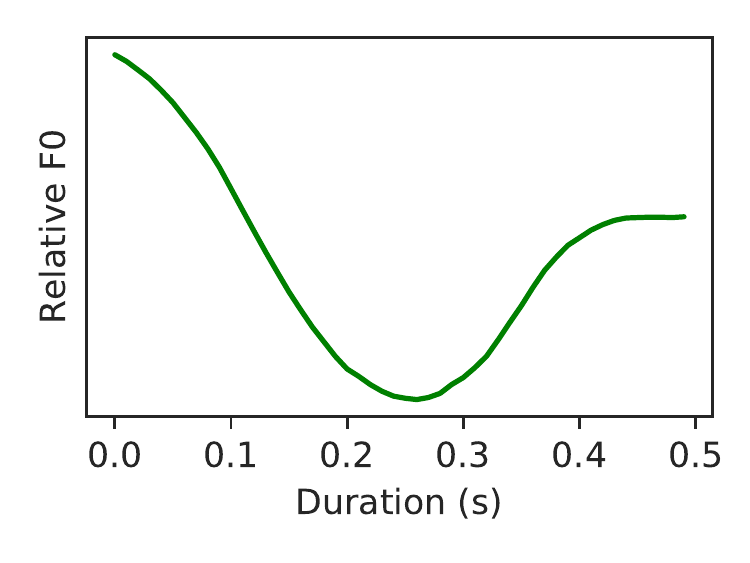}}
        \caption{Incredulity}\medskip
        \label{fig:into_c}
    \end{subfigure}
    \hfill
    \begin{subfigure}[c]{.47\linewidth}
        \centering
        {\includegraphics[width=1.0\linewidth]{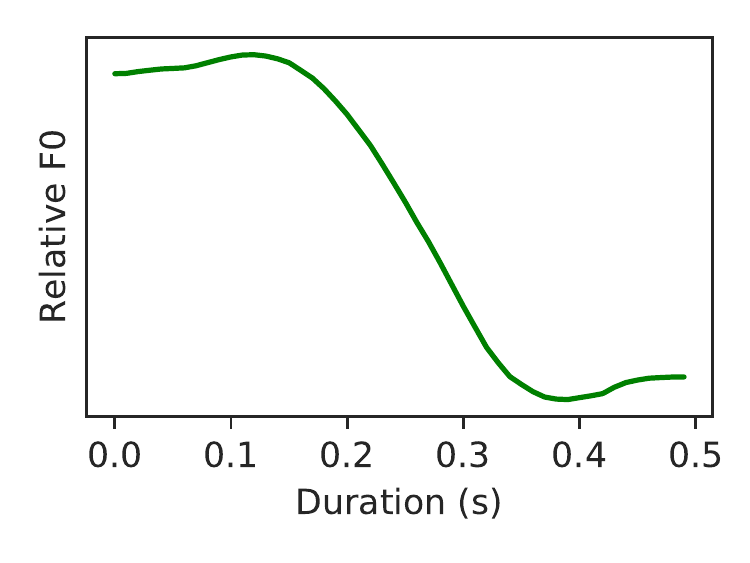}}
        \caption{Obviousness}\medskip
        \label{fig:into_d}
    \end{subfigure}
    \caption{A set of intonation templates that roughly corresponds to the Liberman's classification of English intonation patterns \cite{liberman1975intonational}.
    Each F0 contour represents the centroid of each cluster after k-means clustering.
    }
    \label{fig:into_templates}
\end{figure}

Figure \ref{fig:into_templates} illustrates the F0 contours of the centroids of each intonation cluster after k-means clustering.
We empirically observed that choosing $N = 4$ yields a set of intonation templates closest to the Liberman's classification of English intonation \cite{liberman1975intonational}.
This also has an advantage of providing users with an appropriate number of intuitive options. 
When $N$ is greater than 4, it creates a couple of intonation templates that are similar enough for users without any linguistic background to properly differentiate. 

\subsection{Model Architecture}
\begin{figure}[t]
    \centering
    \includegraphics[width=0.9\linewidth]{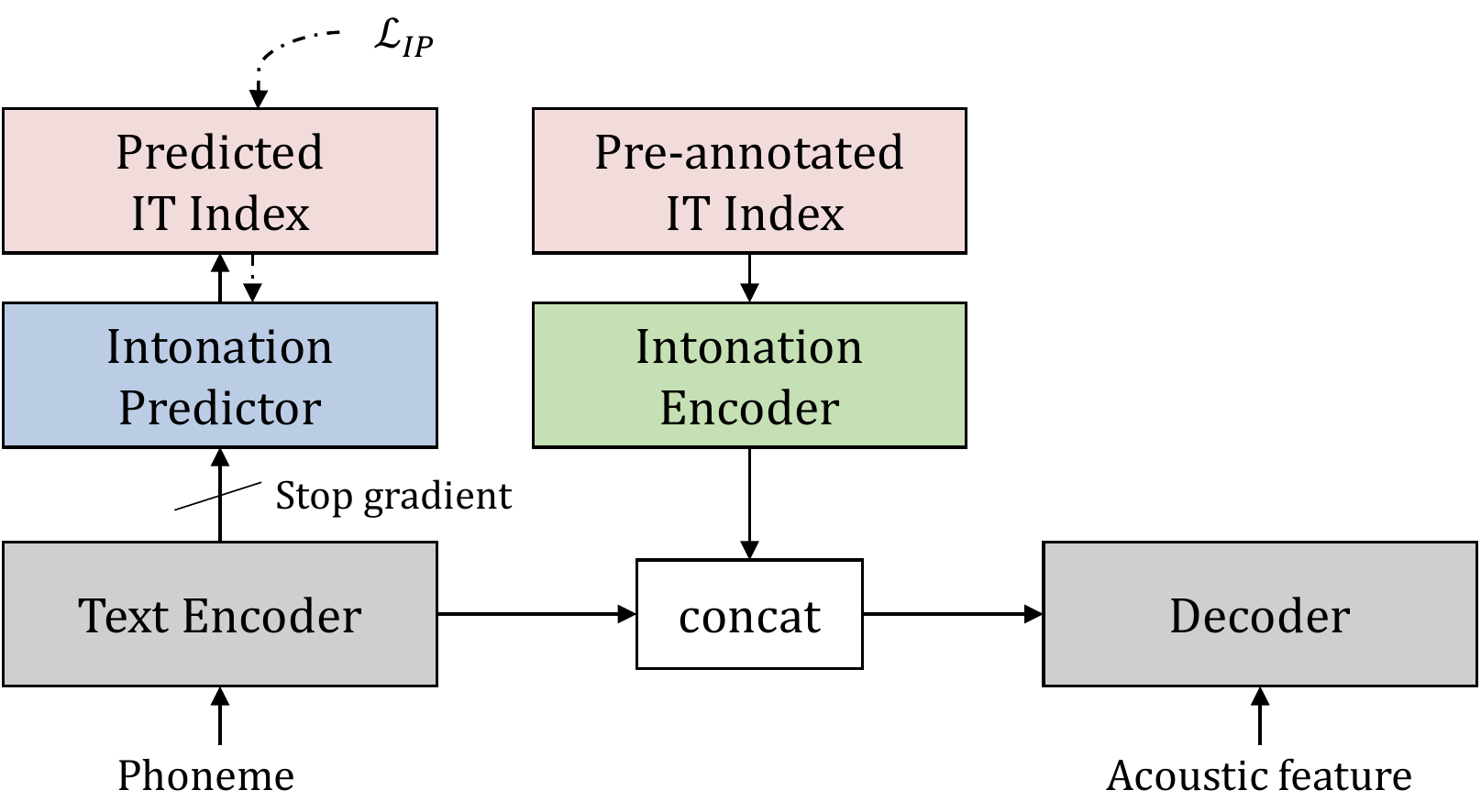}
    \caption{Training scheme. The intonation predictor predicts a suitable Intonation Template (IT) index to the given text. A cross-entropy loss $\mathcal{L}_{IP}$ is utilized to train the intonation predictor. The intonation encoder takes a pre-annotated intonation template index as input, delivering the intonation information to the decoder.
    }
    \label{fig:training}
\end{figure}

Into-TTS is built upon a Tacotron variant \cite{ellinas2020high}.
This Tacotron variant is mainly based on Tacotron 2 \cite{taco2}, with some modifications to utilize the acoustic features from the LPCNet vocoder \cite{valin2019lpcnet}.
We add two modules to this Tacotron variant model: an \textit{intonation predictor} and an \textit{intonation encoder}.
As illustrated in Figure \ref{fig:training}, the intonation predictor learns to recommend a suitable intonation template to the context of text. The intonation encoder handles the intonation information and passes it to the decoder.

\subsubsection{Intonation Predictor}

The intonation predictor takes the text encoder output $\bm{e_{1:l}}$ as input, where $l$ is the length of the input text.
The intonation predictor yields a predicted intonation template index, $\hat{p}$.
The intonation predictor consists of two linear layers.
The first layer has 128 dimensions with ReLU as an activation function.
The second layer has a size of $N$ dimensions, where $N$ is the number of intonation templates. In this paper, we use $N = 4$, as mentioned in Section \ref{into_temp}.
The second layer has Softmax as an activation function.
Finally, an average pooling layer follows, computing the most probable intonation template.
Stop gradient is applied to the input of the intonation predictor to prevent the intonation predictor from affecting any other modules during training.
We add a cross-entropy loss of weight 0.1 to the intonation predictor $\mathcal{L}_{IP}$ along with the loss from  \cite{ellinas2020high} $\mathcal{L}_{Tacotron}$.


\subsubsection{Intonation Encoder}

The intonation encoder takes a pre-annotated intonation template index as input.
It yields the corresponding intonation embedding from the lookup table. Feeding a pre-annotated intonation template index to the intonation encoder helps train the model more efficiently, preventing any errors from the intonation predictor in earlier training steps.
The output from the intonation encoder, $\bm{i}$, is replicated to $\bm{i_{1:l}}$ to match the text length, then is concatenated to the output of the text encoder, providing the intonation information to the decoder.

\subsection{Inference Modes : AUTO and MANUAL}

\begin{figure}[t]
    \centering
    \includegraphics[width=0.9\linewidth]{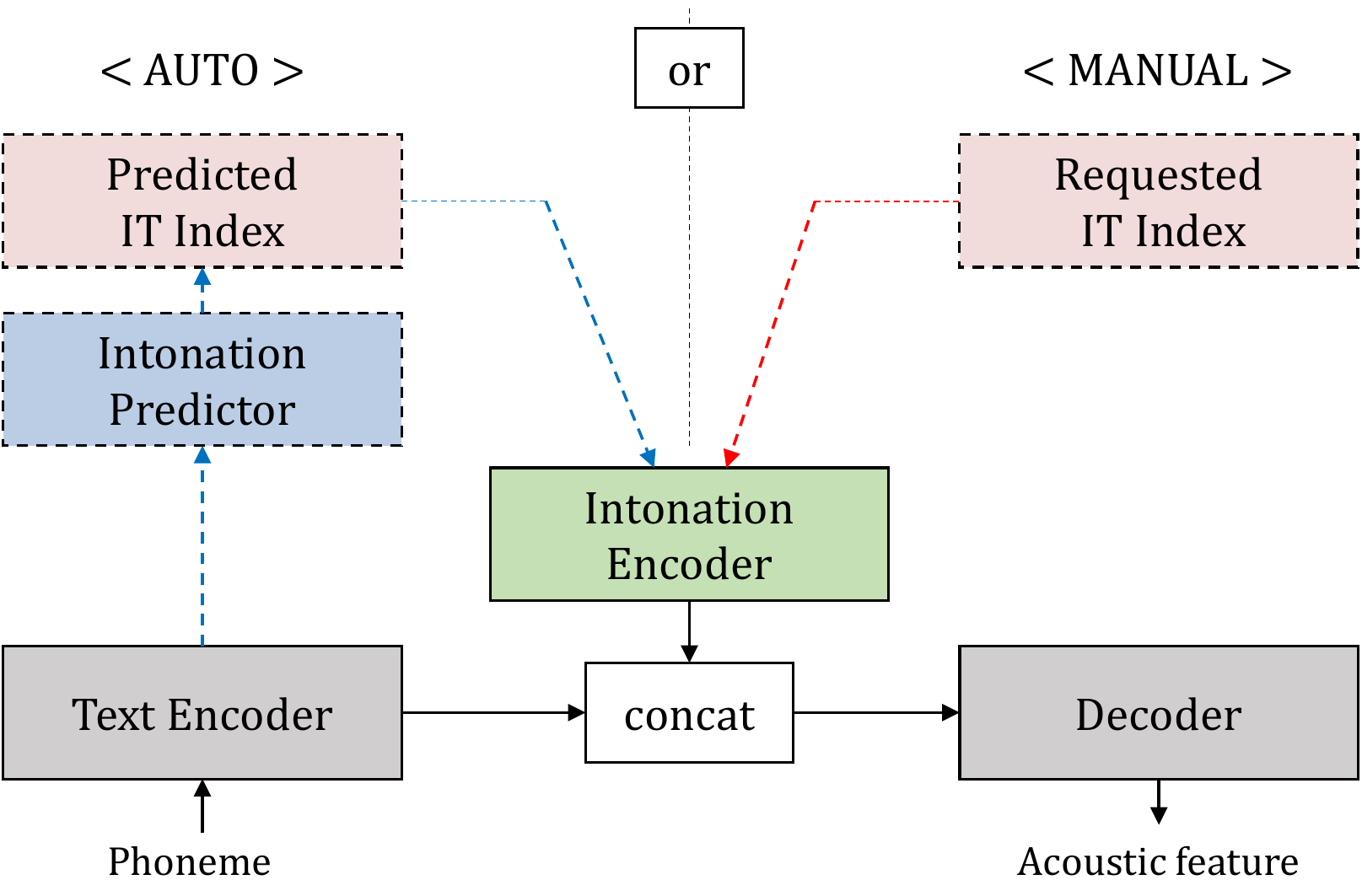}
    \caption{Inference in two modes: AUTO and MANUAL. In the AUTO mode (blue dotted lines), the intonation predictor suggests an Intonation Template (IT). In the MANUAL mode (red dotted line), the requested intonation template is utilized for synthesis.}
    \label{fig:inference}
\end{figure}

During inference, two modes are available: AUTO and MANUAL, as shown in Figure \ref{fig:inference}.
In the AUTO mode, when no specific preference to intonation exists, Into-TTS determines an intonation template automatically.
As illustrated in the blue dotted lines in Figure \ref{fig:inference}, the intonation predictor automatically recommends a compatible intonation template for the context of the input text.
This AUTO mode has a great advantage of making the inference process straightforward by eliminating a mandatory step of determining an intonation pattern, even when no specific preference is given.
Hence, it can cover a wide range of users, even those who are reluctant to choose an intonation pattern for every single sentence.

On the other hand, the MANUAL mode is utilized when an explicit preference to a specific intonation type is given.
In this mode, the intonation predictor is ignored.
Instead, the intonation encoder takes the manually chosen intonation template index as input.
The red dotted line in Figure \ref{fig:inference} represents the inference process in the MANUAL mode.
Even in the MANUAL mode, there is no need to control numerous parameters to control prosody as in the previous reference encoder-based methods.
In users' perspective, they are offered only $N+1$ options: the AUTO mode or one of the $N$ intonation templates.
This simplicity in the inference process lowers the barrier for non-professional users.

\section{Experimental Setup}\label{sec:exp}

An internal database of total 21.2 hours of English speech recordings from 13,000 sentences was utilized for the experiments.
All of the recordings are from two professional voice actors, a male and a female.
300 utterances from each speaker were selected as a test set and excluded from training.
The VCTK dataset \cite{veaux2017superseded} was included along with the internal database to create a set of intonation templates.
We empirically observed that including the VCTK dataset helps generalize intonation patterns among different speakers, yielding a better set of intonation templates.



  



\begin{table*}[th]
  \caption{Pitch distance and Mean Opinion Score (MOS) with 95\% confidence intervals.}
  \label{tab:mos}
  \centering
  \begin{tabular}{l  c c | c | c   }
    \toprule
    \multicolumn{3}{c|}{\textbf{Model Configuration}} & \multicolumn{2}{c}{\textbf{Metrics}} \\
    \midrule
    \multicolumn{1}{c}{\textbf{Model}} & \multicolumn{1}{c}{\textbf{Intonation Predictor}} & \multicolumn{1}{c|}{\textbf{Intonation Encoder}} & \multicolumn{1}{c}{\textbf{Pitch Distance}} & \multicolumn{1}{c}{\textbf{MOS}} \\
    \midrule
    Ground Truth                       &X&X&X&4.46 ($\pm0.06$)\\
    Baseline \cite{ellinas2020high}                     &X&X&X&4.08 ($\pm0.10$)\\
    Baseline-GST &O&X (Reference Encoder)&0.327&4.11 ($\pm0.09$)\\
    \midrule
    \textbf{Into-TTS (proposed)}          &O&O&\multirow{2}{*}{\textbf{0.278}}
    &\textbf{4.18} ($\pm0.09$)\\
    Into-TTS + BERT        &O (BERT)&O&&4.18 ($\pm0.08$)\\
    \bottomrule
  \end{tabular}
  
\end{table*}

The Tacotron variant model \cite{ellinas2020high} was tested as a very \textit{baseline}.
\textit{Baseline-GST} model was compared to investigate if the GST-based approach \cite{wang2018style} can synthesize speech in a specifically requested intonation.
In the baseline-GST model, the reference encoder with the GST is utilized instead of the intonation encoder. In the inference phase, to generate speech in an intended intonation template, we select one representative speech from the training set for each intonation template and feed it to the reference encoder.
The intonation predictor is also added to the baseline-GST model to utilize the AUTO mode as in Into-TTS.
All of the TTS models were trained for 350k steps. 
The bunched LPCNet \cite{Vipperla2020} was utilized as a neural vocoder to convert the acoustic features into actual waveforms.

With recent developments in incorporating pre-trained language model to TTS systems\cite{lei2022msemotts, yoon22b_interspeech, bai22c_interspeech}, a stand-alone intonation predictor finetuned from a pre-trained \textit{BERT\textsubscript{BASE}} model \cite{devlin2018bert} was tested for comparison, in addition to the proposed intonation predictor.
This \textit{BERT predictor} is finetuned for three epochs and takes normalized text as input.

\section{Results and Discussion}


\subsection{Objective Evaluation}

We use the mean pitch distance as the objective evaluation measurement. The pitch distance is calculated as a Euclidean distance of the F0 values of the sentence-final 0.5 seconds between speech samples synthesized by a specific intonation template and the centroid of the corresponding intonation template.
The speech samples from Into-TTS were synthesized in the MANUAL mode, producing four different types of intonations for each sentence.
The corresponding speech samples from the baseline-GST model were synthesized by feeding the speech with smallest pitch distance to the centroid of the corresponding intonation template, to the reference encoder.
The proposed method shows the smallest mean pitch distance for both sets, compared to the baseline GST model as shown in Table \ref{tab:mos}.
The smaller mean pitch distance observed in Into-TTS proves that the proposed intonation encoder synthesizes speech samples in the desired F0 contour more properly, closer to the requested intonation template.

\subsection{Subjective Evaluation}

  

For subjective evaluation, 195 testers were recruited via Amazon Mechanical Turk\footnote{ https://www.mturk.com} for mean opinion score evaluation.
Each tester was provided with randomly chosen 72 speech samples.
The testers were instructed to evaluate the speech samples in the following scale: \texttt{Excellent(5)}, \texttt{Good(4)}, \texttt{Fair(3)}, \texttt{Poor(2)}, and \texttt{Bad(1)} with an interval of 0.5.

The speech samples from Into-TTS and the baseline-GST model were synthesized in the AUTO mode, the intonation predictor automatically choosing an intonation type for each sentence.
As shown in Table \ref{tab:mos}, Into-TTS achieves the highest mean opinion score, outperforming all of the baselines.
The improved mean opinion score of the baseline-GST model suggests that the intonation predictor combined with the reference encoder helps enhance overall naturalness of synthesized speech.
The enhancement in naturalness is even more apparent in the proposed Into-TTS, where the proposed intonation predictor is incorporated with the proposed intonation encoder, as shown in the highest mean opinion score of Into-TTS.
This result proves that our proposed model realizes improved naturalness compared to the previous approaches, even offering an easy-to-use TTS system to users at the same time.
Additionally, similar performance was observed when the intonation predictor had been replaced by the BERT predictor.
This suggests that our proposed intonation predictor's performance is comparable to that of the pre-trained language model.

\section{Conclusion and Future Work}
\label{sec:conclusion}

In this work, we propose \textit{Into-TTS}, an intuitive TTS system to synthesize speech in a desired intonation.
After creating intonation templates in an unsupervised clustering method, we introduce the following two modules to the end-to-end TTS model: the intonation predictor and the intonation encoder.
The template-based approach along with the intonation predictor can cover a wide range of users, from those without any specific preference to those with a specific request.
This method can even allow users with shallow expertise to utilize the TTS system conveniently.
The proposed method outperforms the baseline models in two metrics: pitch distance and mean opinion score.
These results prove that Into-TTS can synthesize speech that follows the desired pitch contour better than the previous approaches, even enhancing naturalness at the same time.

\clearpage
\pagebreak
\ninept



\bibliographystyle{IEEEtrans}
\bibliography{refs}

\begin{thebibliography}{10}
\providecommand{\url}[1]{#1}
\csname url@samestyle\endcsname
\providecommand{\newblock}{\relax}
\providecommand{\bibinfo}[2]{#2}
\providecommand{\BIBentrySTDinterwordspacing}{\spaceskip=0pt\relax}
\providecommand{\BIBentryALTinterwordstretchfactor}{4}
\providecommand{\BIBentryALTinterwordspacing}{\spaceskip=\fontdimen2\font plus
\BIBentryALTinterwordstretchfactor\fontdimen3\font minus
  \fontdimen4\font\relax}
\providecommand{\BIBforeignlanguage}[2]{{%
\expandafter\ifx\csname l@#1\endcsname\relax
\typeout{** WARNING: IEEEtran.bst: No hyphenation pattern has been}%
\typeout{** loaded for the language `#1'. Using the pattern for}%
\typeout{** the default language instead.}%
\else
\language=\csname l@#1\endcsname
\fi
#2}}
\providecommand{\BIBdecl}{\relax}
\BIBdecl

\bibitem{hudson1975meaning}
R.~A. Hudson, ``The meaning of questions,'' \emph{Language}, pp. 1--31, 1975.

\bibitem{hedberg2017meaning}
N.~Hedberg, J.~M. Sosa, and E.~G{\"o}rg{\"u}l{\"u}, ``The meaning of intonation
  in yes-no questions in american english: A corpus study,'' \emph{Corpus
  Linguistics and Linguistic Theory}, vol.~13, no.~2, pp. 321--368, 2017.

\bibitem{skerry2018towards}
R.~Skerry-Ryan, E.~Battenberg, Y.~Xiao, Y.~Wang, D.~Stanton, J.~Shor, R.~Weiss,
  R.~Clark, and R.~A. Saurous, ``Towards end-to-end prosody transfer for
  expressive speech synthesis with tacotron,'' in \emph{international
  conference on machine learning}.\hskip 1em plus 0.5em minus 0.4em\relax PMLR,
  2018, pp. 4693--4702.

\bibitem{lee2019robust}
Y.~Lee and T.~Kim, ``Robust and fine-grained prosody control of end-to-end
  speech synthesis,'' in \emph{ICASSP 2019-2019 IEEE International Conference
  on Acoustics, Speech and Signal Processing (ICASSP)}.\hskip 1em plus 0.5em
  minus 0.4em\relax IEEE, 2019, pp. 5911--5915.

\bibitem{wei2019ICLR}
W.-N. Hsu, Y.~Zhang, R.~Weiss, H.~Zen, Y.~Wu, Y.~Wang, Y.~Cao, Y.~Jia, Z.~Chen,
  J.~Shen, P.~Nguyen, and R.~Pang, ``Hierarchical generative modeling for
  controllable speech synthesis,'' in \emph{International Conference on
  Learning Representations}, 2019.

\bibitem{wei2019ICASSP}
W.-N. Hsu, Y.~Zhang, R.~J. Weiss, Y.-A. Chung, Y.~Wang, Y.~Wu, and J.~Glass,
  ``Disentangling correlated speaker and noise for speech synthesis via data
  augmentation and adversarial factorization,'' in \emph{ICASSP 2019 - 2019
  IEEE International Conference on Acoustics, Speech and Signal Processing
  (ICASSP)}, 2019, pp. 5901--5905.

\bibitem{battenberg2020effective}
E.~Battenberg, S.~Mariooryad, D.~Stanton, R.~Skerry-Ryan, M.~Shannon, D.~Kao,
  and T.~Bagby, ``Effective use of variational embedding capacity in expressive
  end-to-end speech synthesis,'' 2020.

\bibitem{wang2018style}
Y.~Wang, D.~Stanton, Y.~Zhang, R.-S. Ryan, E.~Battenberg, J.~Shor, Y.~Xiao,
  Y.~Jia, F.~Ren, and R.~A. Saurous, ``Style tokens: Unsupervised style
  modeling, control and transfer in end-to-end speech synthesis,'' in
  \emph{International Conference on Machine Learning}.\hskip 1em plus 0.5em
  minus 0.4em\relax PMLR, 2018, pp. 5180--5189.

\bibitem{ronanki2016template}
S.~Ronanki, G.~E. Henter, Z.~Wu, and S.~King, ``A template-based approach for
  speech synthesis intonation generation using lstms.'' in \emph{INTERSPEECH},
  2016, pp. 2463--2467.

\bibitem{vioni2021prosodic}
A.~Vioni, M.~Christidou, N.~Ellinas, G.~Vamvoukakis, P.~Kakoulidis, T.~Kim,
  J.~S. Sung, H.~Park, A.~Chalamandaris, and P.~Tsiakoulis, ``Prosodic
  clustering for phoneme-level prosody control in end-to-end speech
  synthesis,'' in \emph{ICASSP 2021-2021 IEEE International Conference on
  Acoustics, Speech and Signal Processing (ICASSP)}.\hskip 1em plus 0.5em minus
  0.4em\relax IEEE, 2021, pp. 5719--5723.

\bibitem{taco1}
Y.~Wang, R.~Skerry-Ryan, D.~Stanton, Y.~Wu, R.~J. Weiss, N.~Jaitly, Z.~Yang,
  Y.~Xiao, Z.~Chen, S.~Bengio, Q.~Le, Y.~Agiomyrgiannakis, R.~Clark, and R.~A.
  Saurous, ``{Tacotron: Towards End-to-End Speech Synthesis},'' in \emph{Proc.
  Interspeech 2017}, 2017, pp. 4006--4010.

\bibitem{taco2}
J.~Shen, R.~Pang, R.~J. Weiss, M.~Schuster, N.~Jaitly, Z.~Yang, Z.~Chen,
  Y.~Zhang, Y.~Wang, R.~Skerrv-Ryan, R.~A. Saurous, Y.~Agiomvrgiannakis, and
  Y.~Wu, ``Natural tts synthesis by conditioning wavenet on mel spectrogram
  predictions,'' in \emph{2018 IEEE International Conference on Acoustics,
  Speech and Signal Processing (ICASSP)}, 2018, pp. 4779--4783.

\bibitem{ellinas2020high}
N.~Ellinas, G.~Vamvoukakis, K.~Markopoulos, A.~Chalamandaris, G.~Maniati,
  P.~Kakoulidis, S.~Raptis, J.~S. Sung, H.~Park, and P.~Tsiakoulis, ``{High
  Quality Streaming Speech Synthesis with Low, Sentence-Length-Independent
  Latency},'' in \emph{Proc. Interspeech 2020}, 2020, pp. 2022--2026.

\bibitem{liberman1975intonational}
M.~Y. Liberman, ``The intonational system of english.'' Ph.D. dissertation,
  Massachusetts Institute of Technology, 1975.

\bibitem{silverman1992tobi}
K.~E. Silverman, M.~E. Beckman, J.~F. Pitrelli, M.~Ostendorf, C.~W. Wightman,
  P.~Price, J.~B. Pierrehumbert, and J.~Hirschberg, ``Tobi: A standard for
  labeling english prosody.'' in \emph{ICSLP}, vol.~2, 1992, pp. 867--870.

\bibitem{wightman2002tobi}
C.~W. Wightman, ``Tobi or not tobi?'' in \emph{Speech Prosody 2002,
  International Conference}, 2002.

\bibitem{talkin1995robust}
D.~Talkin and W.~B. Kleijn, ``A robust algorithm for pitch tracking (rapt),''
  \emph{Speech coding and synthesis}, vol. 495, p. 518, 1995.

\bibitem{berkovits1984duration}
R.~Berkovits, ``Duration and fundamental frequency in sentence-final
  intonation,'' \emph{Journal of Phonetics}, vol.~12, no.~3, pp. 255--265,
  1984.

\bibitem{valin2019lpcnet}
J.-M. Valin and J.~Skoglund, ``Lpcnet: Improving neural speech synthesis
  through linear prediction,'' in \emph{ICASSP 2019-2019 IEEE International
  Conference on Acoustics, Speech and Signal Processing (ICASSP)}.\hskip 1em
  plus 0.5em minus 0.4em\relax IEEE, 2019, pp. 5891--5895.

\bibitem{veaux2017superseded}
C.~Veaux, J.~Yamagishi, K.~MacDonald \emph{et~al.}, ``Superseded-cstr vctk
  corpus: English multi-speaker corpus for cstr voice cloning toolkit,'' 2017.

\bibitem{Vipperla2020}
R.~Vipperla, S.~Park, K.~Choo, S.~Ishtiaq, K.~Min, S.~Bhattacharya,
  A.~Mehrotra, A.~G.~C. Ramos, and N.~D. Lane, ``{Bunched LPCNet: Vocoder for
  Low-Cost Neural Text-To-Speech Systems},'' in \emph{Proc. Interspeech 2020},
  2020, pp. 3565--3569.

\bibitem{lei2022msemotts}
Y.~Lei, S.~Yang, X.~Wang, and L.~Xie, ``Msemotts: Multi-scale emotion transfer,
  prediction, and control for emotional speech synthesis,'' \emph{IEEE/ACM
  Transactions on Audio, Speech, and Language Processing}, vol.~30, pp.
  853--864, 2022.

\bibitem{yoon22b_interspeech}
H.-W. Yoon, O.~Kwon, H.~Lee, R.~Yamamoto, E.~Song, J.-M. Kim, and M.-J. Hwang,
  ``{Language Model-Based Emotion Prediction Methods for Emotional Speech
  Synthesis Systems},'' in \emph{Proc. Interspeech 2022}, 2022, pp. 4596--4600.

\bibitem{bai22c_interspeech}
Q.~Bai, T.~Ko, and Y.~Zhang, ``{A Study of Modeling Rising Intonation in
  Cantonese Neural Speech Synthesis},'' in \emph{Proc. Interspeech 2022}, 2022,
  pp. 501--505.

\bibitem{devlin2018bert}
\BIBentryALTinterwordspacing
J.~Devlin, M.-W. Chang, K.~Lee, and K.~Toutanova, ``{BERT}: Pre-training of
  deep bidirectional transformers for language understanding,'' in
  \emph{Proceedings of the 2019 Conference of the North {A}merican Chapter of
  the Association for Computational Linguistics: Human Language Technologies,
  Volume 1 (Long and Short Papers)}.\hskip 1em plus 0.5em minus 0.4em\relax
  Minneapolis, Minnesota: Association for Computational Linguistics, Jun. 2019,
  pp. 4171--4186. [Online]. Available: \url{https://aclanthology.org/N19-1423}
\BIBentrySTDinterwordspacing

\end{thebibliography}

\end{document}